\documentstyle{elsart}
\input{epsf.sty}

\begin{document}

\begin{frontmatter}

\title{A WORK- AND DATA SHARING PARALLEL TREE N-BODY CODE}

\author[oact]{U. BECCIANI\thanksref{cnr}},
\author[oact]{V. ANTONUCCIO-DELOGU\thanksref{cnr}} and 
\author[iact]{A. PAGLIARO\thanksref{tracs}}
\address[oact]{Osservatorio Astrofisico di Catania,
Citt\`{a} Universitaria, Viale A. Doria, 6 --
 I-95125 Catania - ITALY}
\address[iact]{Istituto di Astronomia, Universit\`{a} di Catania,
Citt\`{a} Universitaria, Viale A. Doria, 6 --
 I-95125 Catania - ITALY}
\thanks[cnr]{Also: CNR-GNA, Unit\`{a} di Ricerca di Catania}
\thanks[tracs]{Partially supported by the HMC European Community
Program TRACS at the EPCC, UK}
\begin{abstract}
We describe a new parallel N-body code for astrophysical simulations of systems
of point masses interacting via the gravitational interaction. 
The code is based
on a work- and data sharing scheme, and is implemented within the Cray
Research Corporation's ${\rm CRAFT^{\copyright}}$  programming environment. Different data
distribution schemes have been adopted for bodies' and tree's structures. Tests
performed for two different types of initial distributions show that the performance
scales almost ideally as a function of the size of the system and of the number of
processors. We discuss the factors affecting the absolute speedup and how it can be
increased with a better tree's data distribution scheme.

\end{abstract}

\end{frontmatter}

\section{Physical motivation.}

The role of N-body codes as helpful tools of contemporary theoretical cosmology
can be hardly overemphasized. A cursory glance at the specialized astrophysical
literature of the last five years demonstrates that the results of N-body simulations
are often used to check cosmological models, eventually to constrain the free
parameters of these models which cannot be fixed either theoretically or
observationally. Despite their relevance, however, present-day N-body codes can
hardly allow one to deal with more than a few million particles \cite{zqsw94}. Even
using the most simplifying assumptions, we observe in our Universe structures ranging
in mass from the size of a globular cluster ($10^{6} {\rm M}_{\odot}$, where the symbol: 
${\rm M}_{\odot}\approx 1.98\cdot 10^{33} {\rm g}$ denotes the mass of the Sun), up to
clusters and superclusters of galaxies ($10^{15}- 10^{16} {\rm M}_{\odot}$), spanning then a range
of at least $10$ orders of magnitude. Now the ``mass'' resolution of a simulation
of a typical region of the Universe having mass M with ${\rm N}_{\mathrm{p}}$
particles is ${\rm m}={\rm M}/{\rm N}_{\mathrm{p}}$. So, for ${\rm N}_{\mathrm{p}}\leq
10^{7}$ and taking e.g. ${\rm M}=10^{16}{\rm M}_{\odot}$ we have $m\geq 10^{9} 
{\rm M}_{\odot}$, i.e. 3 orders of magnitude larger than the minimum observed mass.
In order to fill this gap it would then be highly desirable to perform simulations with
$10^{9}-10^{10} {\rm M}_{\odot}$. Only codes running on parallel systems can offer, in
some future, the possibility of performing simulations with such a large number of
particles.\\
Among the current algorithms devised to simulate N-body systems of particles
interacting via long-range forces, the one 
based on the \underline{\it oct-tree} decomposition, devised by 
J. Barnes and P.Hut~(\cite{bh86}) bears at least two distinguishing computational
features: it is highly adaptive and its complexity scales as $O(N_{p}\log N_{p})$.
This scaling has been verified in {\it serial} implementations of the 
algorithm \cite{h87}, but it is not at all obvious that the same scaling will
hold for {\it parallel} implementations of the same algorithm. This would be true 
if the additional communication overloads scale as the tree algorithm: which of
course is not true a priori and depends sensitively on the
architecture and on the hardware implementation.\\
Generally speaking, one could expect to meet different problems
when one tries to parallelize the Barnes-Hut algorithm on
Massively Parallel (MP), shared-memory/work systems than on 
Message Passing (MeP) ones. In the latter case a convenient approach consists in 
modifying the original Barnes-Hut algorithm in order 
to parallelize only the most time-consuming part of it
~\cite{s91,ab94,g95}), while in the former case it could prove more rewarding
to exploit 
software and hardware features of
Massively Parallel systems \cite{sw95,s91}. Both approaches have shown 
merits and disadvantages. Starting from the observation that the most
time-consuming part of the serial BH algorithm is tree's traversal analysis,
Salmon \cite{s91} has
introduced a strategy of work sharing based on the introduction of {\it Locally
Essential Tree}s, where each parallel ``task'' builds up a local reduced oct-tree
necessary to compute the evolution of its own particles. This strategy is
based on a {\it spatial} decomposition of the workload among the tasks, and it is easy 
to implement under many popular parallelization environments like
PVM~(\cite{pvm}) and MPI. With this approach the average timestep
execution scales as $T_{step}\propto N^{\alpha}$, with $\alpha$ ranging
from 1.05 to 1.4~\cite{s91,ab94,d96}). This rather large variance depends on {\it
intrinsic} factors (e.g. communication overheads, latency bandwith) and on the
{\it complexity} of the tree. It also affects the critical issue of load balancing.
However on MP systems, it proves possible to avoid these complications and to 
start from the serial BH
algorithm exploiting the available compiler
features on some MP systems which allow the programmer to distribute work and/or
data among the available Processor Elements (hereafter PEs).\\
In this paper we will discuss some of the problems and solutions we have found in 
parallelizing a serial tree code for the Cray's T3D system, within the CRAFT
programing environment. Our solutions can be easily
exported to other MP systems. The issue of {\em portability} is a central one in the
development of {\em any} simulation code. Warren \& Salmon \cite{sw95} have
addressed this problem by developing a small message passing library (called
SWAMPY) which can be
implemented on few platforms, including some workstations, so that their parallel
tree code can be run on some MP as well as on a MeP architecture. We have decided
to follow a more standard strategy, namely to exploit some features of the compilers 
which are available on many MP systems, and to avoid as much as possible data
communication. We then avoid the disadvantage of dealing with a rather
specialized library like
SWAMPY. The disadvantage of our
approach lies in the fact that we have no control on the efficiency of the
Cray's compiler
directives. With the present work we hope to contribute to the understanding of the
tradeoff between these two different programming styles.\\
The code described in this paper was builded starting from a f77 of a serial
Barnes-Hut TREECODE kindly provided to us by Dr. L. Hernquist.
The tests described in this paper were performed on a Cray T3D at the Cineca 
(Casalecchio di Reno (BO) - Italy), a 128 DEC Alpha processor with 8 Mword (64 
bits) memory per processor and 9.6 GigaFLOPS peak performance and on a similar
system with 512 PEs at Edinburgh Parallel Computing Center (EPCC). We have made 
an effort toward portability to other MPP systems by trying to avoid using 
compiler directives which are too specific.  This is why we believe that the 
general strategy outlined in this paper can be easily exported to other MPP 
systems. A similar work has been recently performed on a Fortran 90 
implementation of a direct summation N-body code \cite{sdg94}.\\
In section 2 we describe our parallelization strategy.  In 
section 3 we present our results concerning the scaling and performance. Finally,
in section 4 we report our conclusions.

\section{Parallelization issues}
In our parallel implementation of the Barnes-Hut tree algorithm we have exploited
both the
{\it Data Sharing} and the {\it Work Sharing} programming models.
The flexibility of the CRAFT environment allows one to mix these two modes 
in order to gain the maximum efficiency and speed-up.\\
A detailed description of the Barnes \& Hut parallel algorithm can be found 
elsewhere (\cite{ab94,bh86}). In the following 
paragraphs we will shortly summarize the algorithm main features, and we will 
describe the aspects concerning the parallelizzation issue.
\begin{figure}

\begin{center}
        \epsfxsize=0.4 \textwidth
        \epsfysize=0.7 \textwidth
        \epsfsize={#1 0.8#1}\epsfxsize
        \leavevmode
        \epsffile{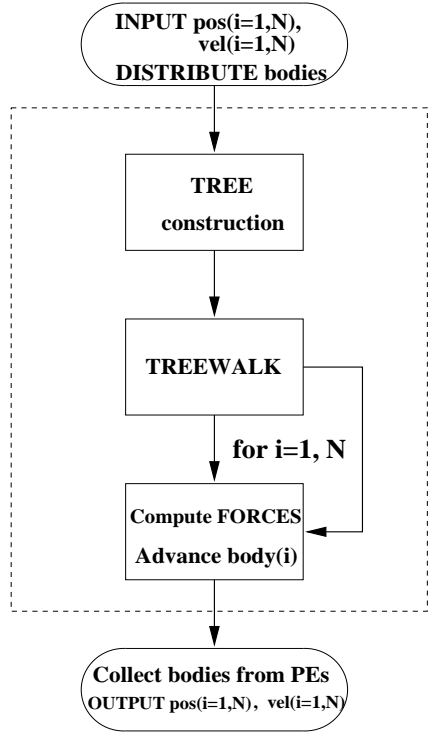}
\caption[]{Block diagram of the Tree N-body code. Steps within the dashed region are
executed in parallel.} \label{fig:1}
\end{center}
\end{figure} 
Apart from the initial and final I/O phases, which cannot be parallelized on the
T3D, 
we can distinguish three phases in our code structure: a) tree formation and cell 
properties calculation; b) 
tree's inspection (treewalk), force evaluation and bodies positions' update; 
c) Dynamic Load Balance (DLB) .\\
In the following paragraphs we will describe these phases.

\subsection{Optimization strategy.}

Data distribution is a very crucial phase to obtain a high 
performance on Cray T3D machine.\\
As a guideline 
we have attempted to share the arrays containing 
the bodies properties among the available PEs in such a way that each PE 
works mostly on bodies resident in its local memory, and at the same time,
to have the same workload for each PE. The CDIR\$ SHARED directive of  CRAFT 
allows data to be shared among all 
available PEs. A data distribution strategy
not correctly tuned  may affect greatly 
the global performance of the code.\\
Generally speaking the execution time of a parallel job 
(${\rm T}_{\rm sol}$) can be 
considered as the sum of a computational time (Tcomp) and of a 
communication time 
(${\rm T}_{\rm comm}$). In turn, the term ${\rm T}_{\rm comp}$
is the sum of an 
operational time ${\rm T}_{\rm flop}$  and data access time 
${\rm T}_{\rm da}$. There are also contribution from the time
needed to redistribute the work (${\rm T}_{\rm dist}$), the time
required to communicate data (${\rm T}_{\rm dd}$) and the time
spent during synchronization (${\rm T}_{\rm sync}$). Eventually we get:
\begin{equation}
{\rm T}_{\rm sol} = {\rm T}_{\rm flop} + {\rm T}_{\rm da} + 
{\rm T}_{\rm dist} + 
{\rm T}_{\rm dd} + {\rm T}_{\rm sync}   \label{eq4}
\end{equation}
The term ${\rm T}_{\rm flop}$ decreases as  the number of PEs involved 
in the parallel run increases; the term ${\rm T}_{\rm da}$ will greatly vary, 
depending on data distribution. 
In our case we have no explicit message passing so the terms ${\rm T}_{\rm
dist}$ and  ${\rm T}_{\rm dd}$ are negligible.
Our goal is that of optimizing the data distribution to allow as many PEs as 
possible to be active during the run, mainly on their 
locally residing data, rather than to work on data located on other PEs, in order to 
minimize the ${\rm T}_{da}$ term.\\
In order to achieve this target we have attacked the problem of data distribution from
two sides: a) optimizing the distribution of bodies among PEs; b) optimizing the
distribution of the tree, i.e. of its cells.


\subsection{Tree formation and cell properties.}

The spatial domain containing 
the system is divided into a set of nested cubic cells by means of an oct-tree 
decomposition.\\
At the beginning all the computational domain is enclosed within a cubic region
called {\it root cell}
containing all the particles, which is divided into 8 subcells. 
This step is repeated for each cell of the tree until one arrives to cells containing 
only 1 body. This structure is the 
\underline{\em tree}. For cells containing more than  one body ({\em internal}
cells, hereafter {\em icells}),
positions, sizes, total 
mass and quadrupole moments are stored in corresponding arrays. For cells 
containing only 
one body ({\em terminal} cells, hereafter {\em fcell}), on the other hand, only 
the position of the cell is stored. Observe that only cells containing {\em at least}
one particle are kept in the tree, so at each new level of the hierarchy (i.e. at each
{\em depth} $d$ of the tree) at most $2^{3d}$ new cells are added.\\ 
Making use of the work sharing model, all the available PEs contribute to tree 
formation and to {\em icells} properties calculation. Parallelism is 
attained by sharing the loops structures among the PEs.  The CRAFT directive  
CDIR\$ DOSHARED (ind1,[ind2, ind3,...]) mechanism  allows one to share a do loop: 
inside the shared loop each PE executes its assigned loop iteration, as in the
following example.\\

\hspace* {1cm} REAL pos(1024,3), pm1(8,3), cellsize(1024)\\
\hspace* {1cm} CDIR\$ SHARED pos(:BLOCK(Np/Npes,:)), cellsize(:BLOCK)\\
\hspace* {2 cm} ndim=3 \\
\hspace* {1cm} CDIR\$ DOSHARED (p,k) ON pos(p,k)\\
\hspace* {2 cm} DO k=1,ndim\\            
\hspace* {2.5 cm} DO p=1,nsubset\\
\hspace* {3 cm} pos(p,ndim)=pos(p,ndim)+pm1(j,ndim)*0.5*cellsize(p)\\
\hspace* {2.5 cm} ENDDO\\
\hspace* {2. cm} ENDDO\\

Each $(k,p)$ iteration is executed only by that PE having in its own local 
memory the pos(p,k) element. However, in order to compute forces on those bodies
residing on some particular PE, segments of the tree residing in some other PEs
need to be accessed. These remote-access operations will result in a slowing down of
the code; the actual amount will depend ultimately on the average (over all
bodies) length of the interaction list.
Using Apprentice, a performance 
analyzer tool designed for CRAY MPP, we have noted that  
about 65\% of the work is performed in parallel by the available PEs. This fraction
tends to increase with increasing number of particles $N_{\rm P}$, for a fixed number of
processing elements, $N_{\rm PE}$, because having more particles allocated to
each processor the `{\em granularity}' (i.e. the
amount of computational work allocated to parallel tasks), tends to increase. This
will also result in a percentually lower communication overhead.

\subsubsection {\it Tree properties: data distribution}

\begin{figure}
\begin{center}
        \epsfxsize=0.7 \textwidth
        \epsfysize=0.7 \textwidth
        \epsfsize={#1 0.5#1}\epsfxsize
        \leavevmode
        \epsffile{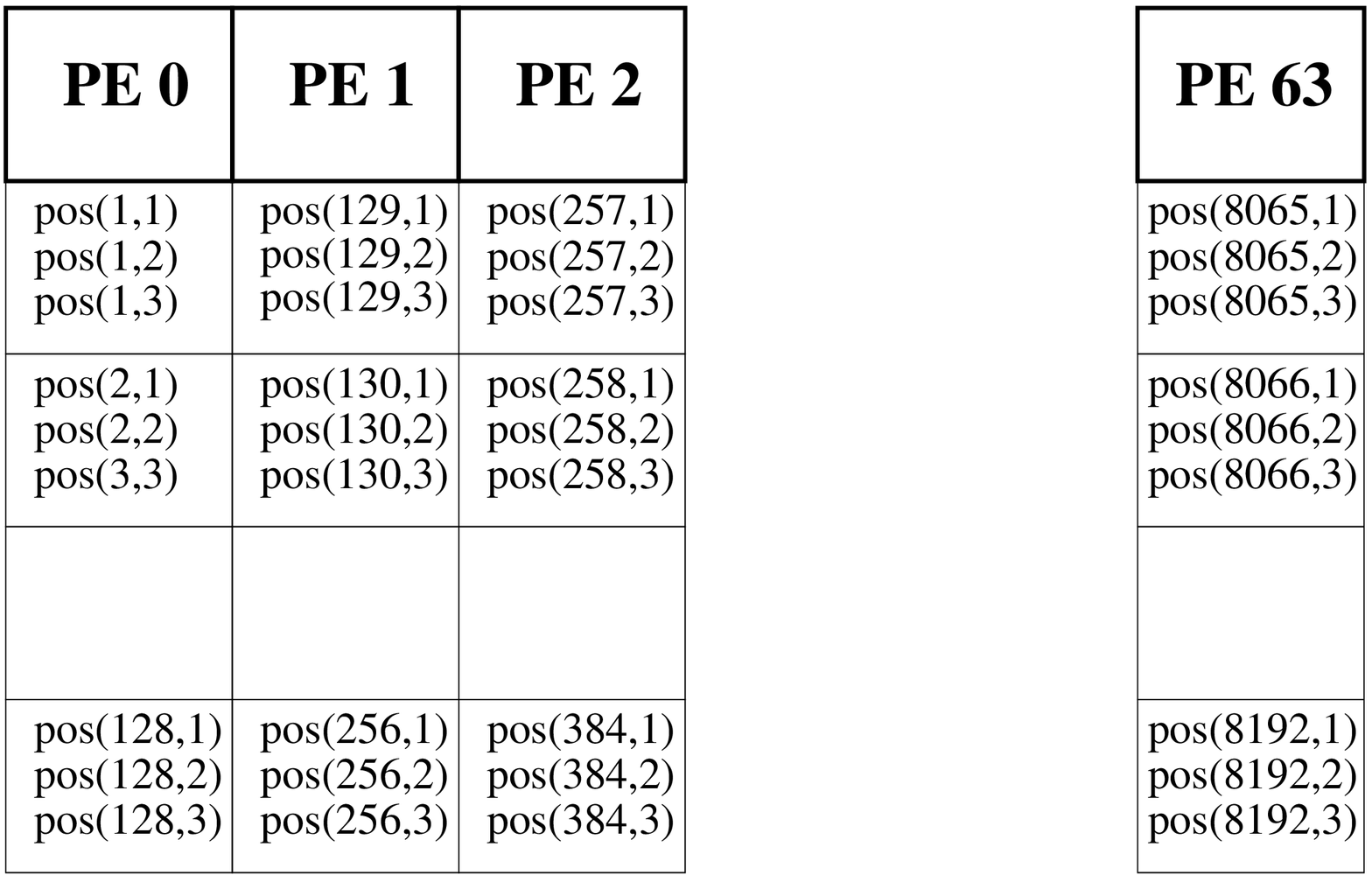}   
\caption[]{Data distribution for 64 PEs, of an array 
pos(8192,3) containig bodies' properties. 
Each PE has residing nearest bodies.} \label{fig:2}
\end{center}
\end{figure}

The data distribution scheme of the arrays containing the tree properties (cells size, 
geometric and physical characteristics) was adopted
after many trials varying the number of bodies (from 1,000 up to 216,000), 
and using several layout of data distribution. The optimal distribution was reached using 
a fine tree data distribution with directives like: CDIR\$ SHARED 
CELLSIZE(:BLOCK,:) as shown in fig.~\ref{fig:3}. Note that this distribution is different 
from
that of the bodies (figure~\ref{fig:2}): contiguous cells here are mostly distributed in arrays
`lay' perpendicularly to the PEs' distribution. The reason why this distribution results
in better performance can be easily understood 
when one considers the cells' {\em spatial} distribution. 
Cells are numbered progressively from the root (which 
encompasses the
whole system)  down to the smallest cells which enclose smaller and smaller regions of
space. The first, say, $2^{9}+1$ cells (i.e. all the cells down to depth $d=3$) are 
typically enough large to contain many bodies,
at least during the initial steps of a cosmological simulation, when the configuration
is almost homogeneous. This means that almost all the bodies will have to inspect the
first cells in the tree's hierarchy: the typical timescale of this process will be
determined by the access time of the bodies residing in the farthest PEs. If {\em all} 
the cells were distributed in a fine grained way, each PE on average
will contain an equal amount of cells at any level of the hierarchy (at least for those
hypercube depths $d_{hyp}$ such that $d_{hyp}\gg \ln_{2}N_{PE}/3$), so that on
average each body will have the same access time to the tree, independently of where
 its parent PE is located.\\
We do not claim that this is the optimal choice for mapping the tree onto the T3D's
torus. The problem of efficiently mapping domains onto an underlying harware
topology is 
stilll a matter of debate (see e.g.~\cite{letal95}).

\subsection{Force calculation and system's update.}

This phase is consuming more than 80\% of the 
total computational time in the serial code. For each body, the calculation 
of the force is made 
through inspection of the {\em tree}, forming an {\it interaction cell list}: 
in particular, one compares the relative position of the cells of the tree 
with that of the body.  Let ${\bf r}_{i}, {\bf r}_{c}^{(k)}$ be the 
position vector of the $i$-th body and of the $k$-th cell, respectively. 
We introduce some ``{\em distance}'' $d({\bf r}_{i}\mid{\bf r}_{c}^{(k)})$ 
between the body and the cell. This could be the distance between the 
body and the center of mass of the given cell. Considering the ratio 
$z_{OC}=l^{(k)}/d({\bf r}_{i}\mid{\bf r}_{c}^{(k)})$ for $1\leq k\leq 
N_{cell}$, where $l^{(k)}$ is the size of the $k$-th cell (assuming that 
$k=1$ is the root cell), those cells for which $\theta< z_{OC}$ (where 
$0\leq\theta\leq 1$) are considered too ``nearby'' to the body, and are not 
added to the interaction cell list. This means that the particle ``sees'' 
these cells as extended objects and one needs to ``look inside'' the smaller 
cells contained within them. For each of these one 
recalculates $z_{OC}$ and one checks whether it is larger or lesser than 
$\theta$:  cells which do not satisfy the criterion and terminal cells (i.e. cells
containing only one body) are 
included in an interaction cell list. The calculation of the force is 
made using this list, and at the end the bodies positions on each body and
velocities are advanced of one time step.
\\  Each PE executes the routines for this 
segment separately in a parallel region code, and mostly works on those bodies 
that are resident in the 
local memory, as described in the data distribution phase (figure~\ref{fig:2}). Using 
this calculation scheme, no specific sychronization mechanisms are needed.
A dynamic 
load balancing is active in this phase and can re-distribute the load between 
PEs. Using Apprentice,  we have noted that 
100\% of the work is performed in parallel by the 
available PEs.
At the end of this phase there is a specific synchronization mechanism 
(CDIR\$ BARRIER), before updating body position and start the next step.\\

\subsubsection{\it Bodies properties: data distribution} 

The arrays containing the bodies properties (position, mass, 
velocity, acceleration) are spread in contiguos block, using the CDIR\$ 
SHARED POS(:BLOCK({\rm Np}/{\rm Npes}),:) directive, as shown in figure~\ref{fig:2}. The initial conditions file, 
containing mass, position and velocity terms of the bodies, must be constructed in 
such a way that bodies near in space are labelled with nearby integers, in order to
increase the probability that all nearby bodies lie on the same or very near PEs.\\
If  necessary a sort of bodies position array and a memory array re-distribution 
may be executed, to preserve these properties during the run.

\begin{figure}
\begin{center}
        \epsfxsize=0.7 \textwidth
        \epsfysize=0.7 \textwidth
        \epsfsize={#1 0.5#1}\epsfxsize
        \leavevmode
        \epsffile{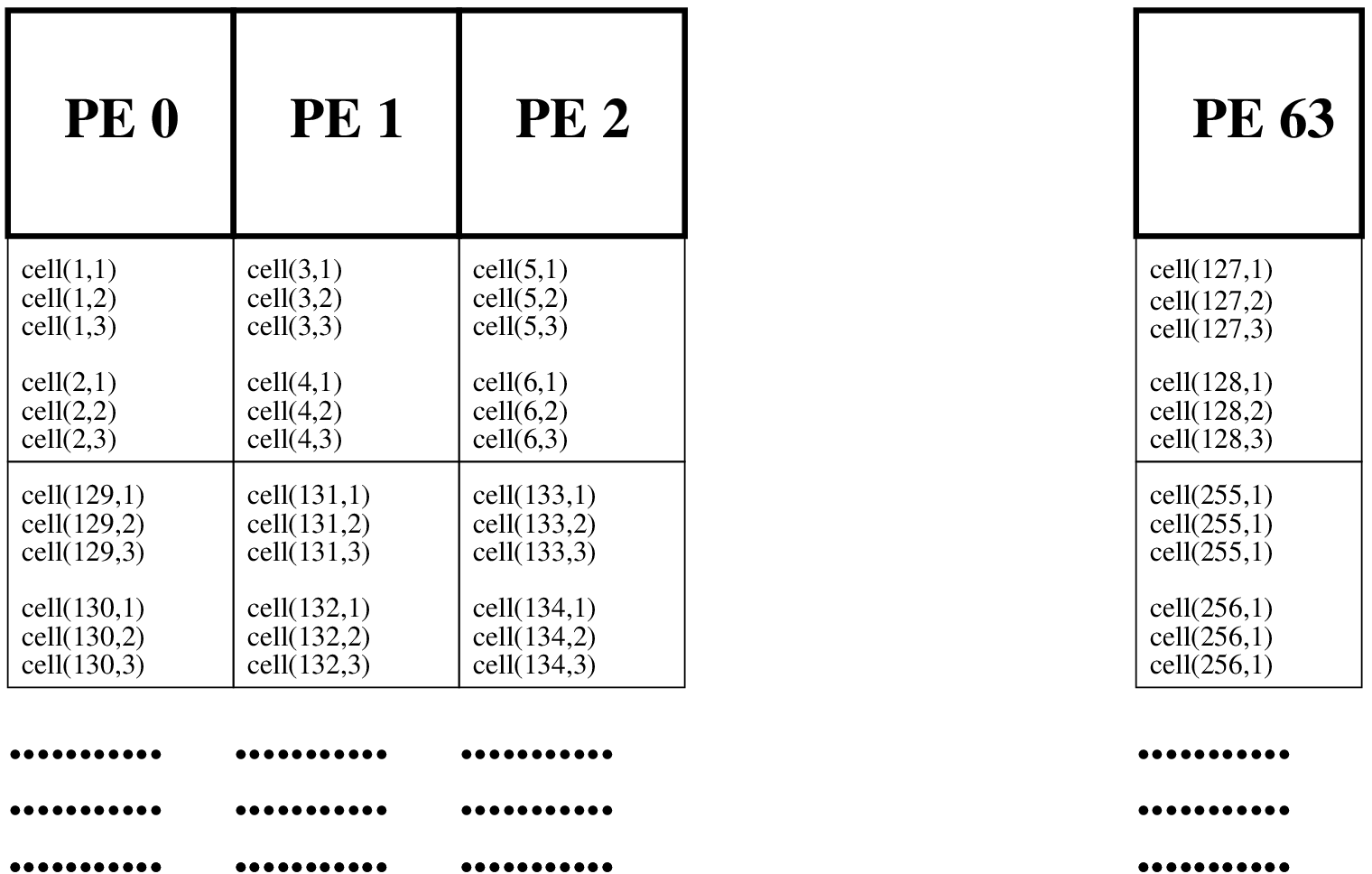}   
\caption[]{Data distribution among 64 PEs of an array containing tree 
properties (cells size, quadrupole moments).} \label{fig:3}
\end{center}
\end{figure}


\subsection{Dynamical Load Balance}

In an ideal situation during the run all PEs would perform the same work 
consuming the same time. A load imbalance arises when one or more PEs, 
most probably during the TREEWALK, spends more time than 
others PEs (up to a fixed threshold); consequently all the code will run at the
speed of the slowest PE, and this will greatly affect 
the total performance of the run.
The workload depends strongly on
bodies' data distribution being {\em uniform} or {\em inhomogeneous}: 
ultimately on the geometry and mass distribution 
of the particles within the system, and can greatly vary during the run
when clusters of particles  form.\\
The Dynamical Load Balance (DLB) routines that we have implemented help
us to avoid that during the run an imbalance situation arises.\\
The DLB structure is based on 
the assignment to each body of a PE executor (PEX(i)): the PE that executes 
the {\em force calculation} phase for the body. At the beginning, 
using the data distribution shown in figure~\ref{fig:3}, the PEX assigned to 
each body is the PE where the body properties are residing.\\
When the bodies' distribution evolves toward inhomogenous, clustered
configurations, some PEs begin to consume a very high time to
 evaluate the forces on the local bodies in comparison with other PEs,
and a load redistribution is performed at the end of each time-step by means of
 the following scheme.

\subsubsection{\em Workload estimation.}

Assuming to have K availables processors ${\rm PE^{(L=1,...K)}}$, each having 
$N_{\rm K}$ residing bodies, the load of each processor  is evaluated as: 

\begin{equation}
W^{\rm (L)}_{\rm LD} = \sum_{\rm i=1}^{N_{\rm L}} BDLD^{\rm (L,i)}
\end{equation}
where we have introduced the per-particle workload $BDLD^{\rm (L,i)}$  defined as 
the time spent to evaluate the force acting on the i-th body assigned to the L-{\em th}
 PE (in cpu clock cycles). The average load after each timestep is then:
\begin{equation}
AVLD = \frac{\sum_{\rm L=1}^{\rm K} W^{\rm (L)}_{\rm LD}}{\rm K}
\end{equation}

\subsubsection{\em PEX(i) assignment.}

\begin{figure}
\begin{center}
        \epsfxsize=0.7 \textwidth
        \epsfysize=0.7 \textwidth
        \epsfsize={#1 0.5#1}\epsfxsize
        \leavevmode
        \epsffile{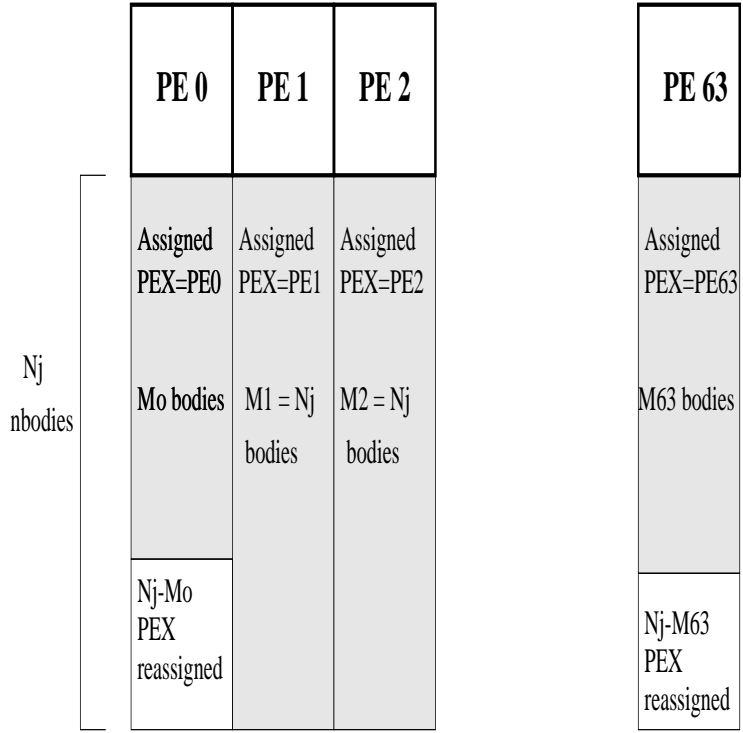}   
\caption[]{Dynamical Load Balance. PE executors assignment.} \label{fig:4}
\end{center}
\end{figure}

Considering the $N_{\rm K}$ bodies residing on $PE^{(\rm L)}$'s local memory
an integer pointer ${\rm PEX=}PE^{\rm (j)}$ to the processor is assigned to the first 
$M_{\rm k}$ bodies ($M_{\rm k}\leq N_{\rm k}$), where  $M_{\rm k}$ is determined 
in such a way to fulfill the condition:
\begin{equation}
\sum_{\rm i=1}^{\rm M_{K -1}} BDLD^{\rm (L,i)} < AVLD \leq 
\sum_{\rm i=1}^{\rm M_{K}}BDLD^{\rm (L,i)}
\end{equation}
(see figure~\ref{fig:4}). After these assignments, each PE has an estimated load not very different from
 AVLD . There could still be $N_{\rm K} - M_{\rm K}$ bodies which have not been assigned to
any PE other than that on which they are residing.

\subsubsection{\em PEX reassignment.}

During this phase each PE assigns a PEX to the $N_{\rm k} - M_{\rm k}$ 
remaining bodies (if any).
Each PE follows this rule for the PEX assignment: for each of the remaining bodies 
all the processors (including the processing one) are put on a list, and the PEX
is assigned to that PE having the smaller load.\\
In this way it turns out to be possible to minimize the $T_{\rm sol}$ in eq.~\ref{eq4} 
minimizing the $T_{\rm da}$ term.
The use of this technique of Load Balancing 
results in a gain from 10\% up to 25\%.

\subsection{Memory Requirements.}
The memory requirements of the Tree N-body codes are generally large due to
the presence of arrays containing the particles and the tree cells properties.
In a Locally-Essential-Tree-based code, the larger part of the memory occupancy
is due to the  dimension of arrays containing the local particles' and tree
cells properties. Moreover a relevant part of memory occupancy must be used for
the additional arrays containing particles and cells imported from other
processors. This quantity increases with decreasing $\theta$ and is not {\em
a priori} predictable, so that all the arrays must be dimensioned for the worst case.\\
In our work-memory/shared code the memory occupancy is lesser, because the
absence of Locally Essential Trees reduces the replication of arrays.
Let $N_{\rm bodies}$ be the number of particles.  The memory occupancy of 
our code is given by:
\begin{equation}
{\rm M}=\frac{40N_{\rm bodies}}{N_{\rm PE}}+5\frac{N_{\rm bodies}}{C}
\end{equation}
words (1 word=8 bytes) for each PE. In the above equation $N_{\rm PE}$ is the 
number of PEs involved in the simulation and C a factor
depending on the  length of the interaction list formed by each particles
during the tree traversal phase. An upper minimum value for this latter quantity
is N/10. Using this value of C and for a total number of 256 PEs , 
it is possible to run a simulation with about $N_{\rm bodies}\approx 10^{6}$ on
resources like those offered by the CINECA T3D (128 PEs, 2 Gbytes RAM).
In the nearest future, using the T3E machine with 16 Mword 
for each node, more than 20 million particles simulation may be run.
These figures may increase for $C>10$, and this fact opens the possibility of
performimg simulations with 30-40 million particles even on MPP system having a
moderate amount of mass memory e.g. of the order of a few distributed Gbytes.

\section{Results}
In order to compare the performance of our code in realistic situations we have run
tests for both homogeneous and inhomogeneous initial conditions .
At variance with the Molecular Dynamics case, the gravitational
force induces an irreversible evolution towards highly clustered configurations. From
the computational point of view this results into load imbalance, so a comparison of
the two cases provides information on the efficiency of the 
Dynamical Load Balance
procedure described before.\\
The central subroutine STEPSYS which advances the system's 
positions and
velocities of one time step can be schematically decomposed 
into three main phases: a) MAKETREE,
where the oct-tree is builded; b) ACCGRAV, in which {\em each} 
particle inspects the
tree, builds up an interaction list, and the acceleration on each
particle is computed; c) STEPPOS and STEPVEL, where the system 
is advanced. This
latter step is executed in parallel by each PE.\\
Following Warren and Salmon~\cite{ws95}, we present our results plotting
the quantity:
\begin{equation}
T_{\rm speed}=\frac{T_{\rm step}N_{\rm PE}}{N_{\rm bodies}}
\end{equation}
 where $T_{\rm step}$ is the
{\em normalized} average execution time of STEPSYS over 10 time steps (measured
w.r.t. the serial case). In an ideal case, one would expect that 
$T_{\rm step}\propto N_{\rm PE}^{-1}$. For the serial code one also expects: 
$T_{\rm step}\propto N_{bodies}logN_{bodies}$. 
\begin{figure}
\begin{center}
        \epsfxsize=0.7 \textwidth
        \epsfysize=0.7 \textwidth
        \epsfsize={#1 0.5#1}\epsfxsize
        \leavevmode
        \epsffile{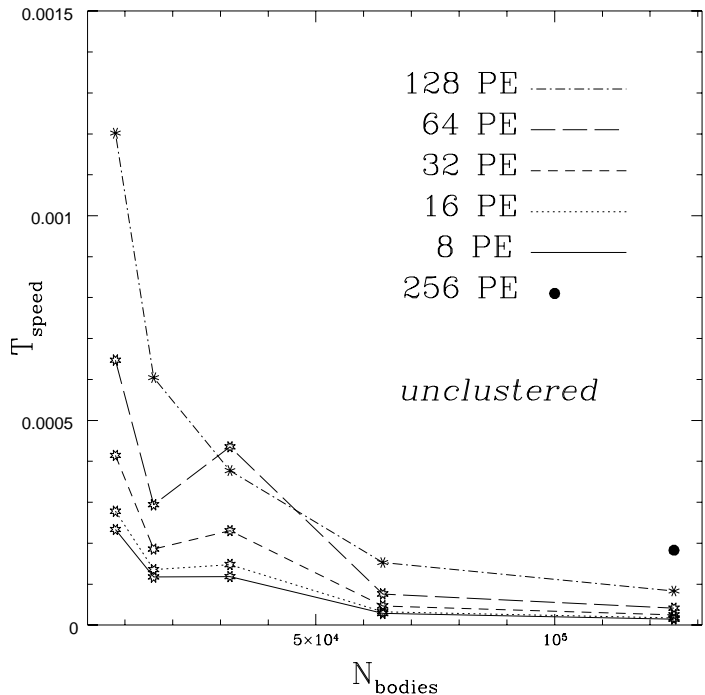}   
\caption[]{Scalability of the relative speedup for different partitions. $T_{step}$
is normalized to the value measured for 1 PE.} \label{fig:5}
\end{center}
\end{figure}
As one can see from Fig.~\ref{fig:5}, the code scales very efficiently for large
number of particles: $T_{\rm speed}$ reaches an asymptotic regime already at
$N_{\rm bodies}\approx 2\times 10^{5}$. It is interesting to observe that the
asymptotic regime is reached also for high granularity cases, i.e. for $N_{\rm PE}=8,
16$. The same trend concerning the scalability with $N_{\rm bodies}$ is observed for
the scalability with $N_{\rm PE}$ (Figure~\ref{fig:6}). The anomalous behaviour of the
run at 32k, visible also as a `shoulder' at the corresponding point in
Fig.~\ref{fig:5}, is
due to spurious cache effects.

\begin{figure}
\begin{center}
        \epsfxsize=0.7 \textwidth
        \epsfysize=0.7 \textwidth
        \epsfsize={#1 0.5#1}\epsfxsize
        \leavevmode
        \epsffile{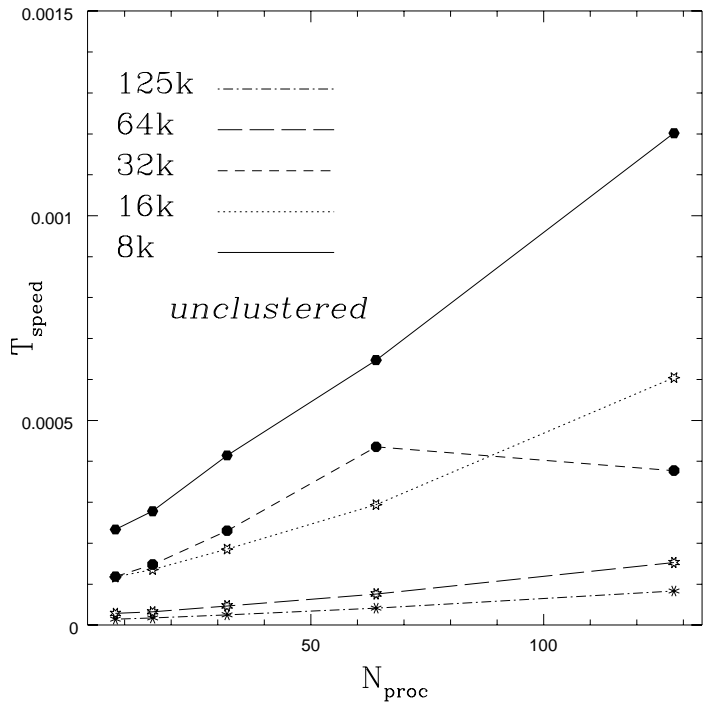}   
\caption[]{Scalability of the relative speedup for different 
sizes of the 
running code. $T_{step}$ is normalized as in Figure~\ref{fig:5}}
 \label{fig:6}
\end{center}
\end{figure}
\begin{figure}
\begin{center}
        \epsfxsize=0.7 \textwidth
        \epsfysize=0.7 \textwidth
        \epsfsize={#1 0.5#1}\epsfxsize
        \leavevmode
        \epsffile{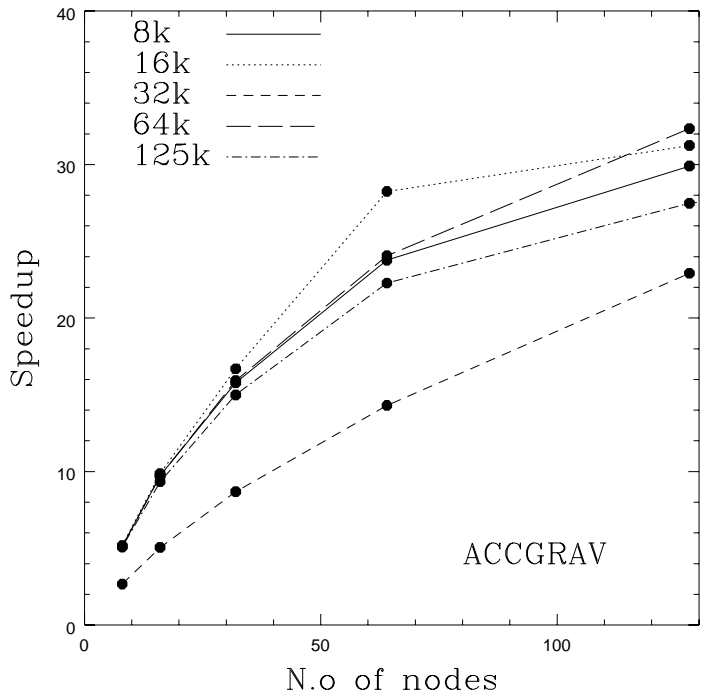}   
\caption[]{Speedup for ACCGRAV.} \label{fig:7}
\end{center}
\end{figure}

In Figure~\ref{fig:6} we plot the relative speedup
$T_{\rm speed}$ as a function of $N_{\rm PE}$ for homogeneous inital conditions.\\ 
Note that $T_{\rm speed}$ provides a measure of the {\em scalability} of the code,
not of its absolute speedup. This latter is plotted in Figure~\ref{fig:7} for the
subroutine ACCGRAV. Note that measuring ACCGRAV we are in fact
measuring the performance of TREEWALK, the most time-consuming subroutine and
the one which is fully parallelized. But the performance is also
influenced by other factors, like those mentioned in section 2.1, which
altogether act to reduce the absolute speedup. We believe however that
it would be prove possible to increase futher the speedup with a
dynamical tree allocation, and we are working along this direction.
It is interesting to observe that our results for $T_{\rm
speed}$ are comparable to 
those obtained by Warren and Salmon~\cite{ws95} (their Fig. 8). We cannot say very
much about their absolute speedups, because they do not give any information about
it.

\begin{figure}
\begin{center}
        \epsfxsize=0.7 \textwidth
        \epsfysize=0.7 \textwidth
        \epsfsize={#1 0.5#1}\epsfxsize
        \leavevmode
        \epsffile{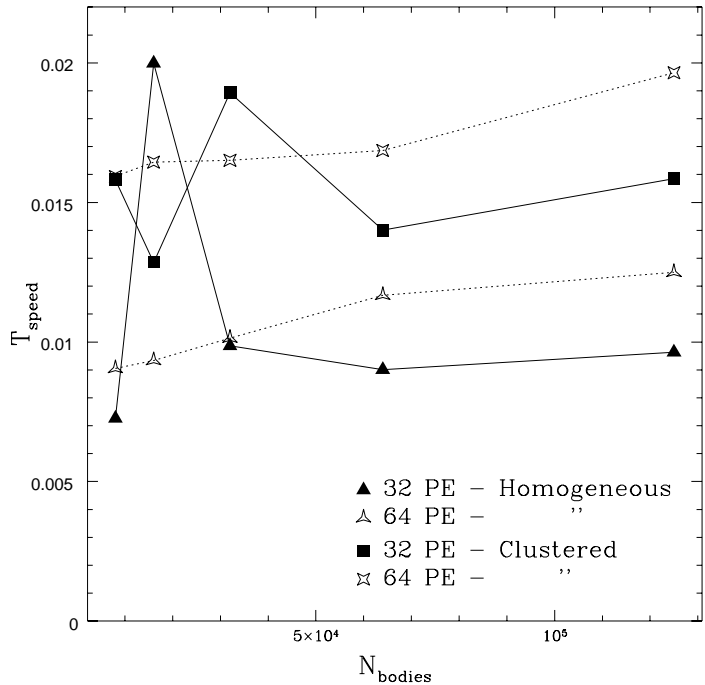}   
\caption[]{Performance degradation with increasing clustering. } \label{fig:8}
\end{center}
\end{figure}

As we mentioned at the beginning, Load Balancing is also necessary in order to avoid
the performance degradation one meets when the system starts to cluster. In
figure~\ref{fig:8} we can appreciate how much this problem quantitatively affects . The
quantity $T_{\rm step}$ almost doubles when one passes to clustered configurations,
and also the steepness of the curves tends slightly to increase, although not
dramatically. This means that the scaling properties of the codes keep almost
unchanged with increasing clustering, and that even for highly inhomogeneous
configurations there is not a significant increase of the communication overhead
among the main sources of load unbalance.

\section{Conclusions}
The work- and memory-shared Tree N-body code we have described in this
paper has some very interesting features. First, its memory occupancy is
comparatively lesser than in a LET-based scheme, because in this latter some parts of
the local trees have to be replicated on other PEs. This reduced memory occupancy
results also in a reduced communication overhead, simply because the structures
relevant for the force calculation are already shared and they have not to be
exchanged as in the LET, message passing schemes. Being based on a different
algorithm, the scalability of our
code cannot be {\em a priori} assumed to be the same as for explicitly
message-passing implementations as those developed by many
authors~\cite{s91,ab94,sw95,ws95,d96}.

 We observe however that as  far as {\em
scalability} (measured by $T_{\rm speed}$) is
concerned, our code performs very well, in a way very similar to that observed in
LET-based implementations. We think that a dynamical tree
allocation, i.e. a scheme in which the block distribution of
the tree changes with time, could increase the absolute speedup. 
The atomic, almost uniform distribution
scheme we have adopted was motivated by the very good scalability, but in very
inhomogeneous situations it could happen that a body residing in a given PE needs
to access data from `deep' cells lying on some far PE. Due to synchronization
mechanisms this can occasionally result in a general slowing down of the code.\\
Our main purpose in this paper was to try to understand with specific tests these
problems and how they affect the relative performance of different parts of the code.
In a forthcoming paper~\cite{abep96} we will discuss a Load Balancing scheme based
on a dynamic data sharing distribution scheme.
\begin{ack}
The tests quoted in this paper were performed on the Cray T3Ds at CINECA,
Casalecchio di Reno (Bologna), ITALY and at Edinburgh Parallel Computing Center,
Edinburgh, UNITED KINGDOM. We have benefitted from the support from prof. G.
Erbacci (CINECA). AP benefitted of a TRACS EC grant at EPCC,
and wishes to thank S. Paton (Cray Research Corp.) for some
helpful discussions during his stay in Edinburgh. 
\end{ack}
{}


\begin{thebibliography}{}

\bibitem{abep96}
V. Antonuccio-Delogu, U. Becciani, G. Erbacci and A. Pagliaro, {\it in
preparation} (1996)

\bibitem{ab94}
Antonuccio-Delogu, V. and Becciani, U., "A Parallel Tree 
N-Body Code for Heterogeneous Clusters", in: J. Dongarra and J. Wasniewsky, 
eds.,{\it Parallel Sciebtific Computing - PARA '94} (Springer Verlag: 1994), 17

\bibitem{bh86}
J. Barnes and P. Hut, {\it Nature}~{\bfseries 324} (1986) 446

\bibitem{cr94}
Cray Research Inc., {\it "Cray MPP Fortran Refeference Manual} 
SR-2504 6.1 (1994)

\bibitem{d96}
J. Dubinski, {\it "A Parallel Tree Code"}, 
submitted to {\it New Astronomy} (1996)

\bibitem{g95}
Gouhong Xu, {\it "A new parallel N-body gravity solver: TPM"}, 
{\it Astrophys. J. Supp.} {\bfseries 97} (1995) 884

\bibitem{pvm}A. Geist, A. Beguelin, J. Dongarra, W. Jiang, 
R. Manchek and V. Sunderam, "{\it PVM 3 User's Guide and Reference 
Manual}", ORNL/TM-12187, September 1994

\bibitem{letal95}
V. Lamsani, L. Bhuyan and  D. Scott Linthicum, {\it Parallel Computing} {\bf 21} (1995)
993
\bibitem{h87}
L. Hernquist, {\it Astrophys. J. Suppl.} {\bf 64} (1987) 715

\bibitem{s91}
J.K. Salmon, {\it "Parallel hierarchical N-body methods"}, Ph. D.. Thesis, 
{\it unpublished } (California Institute of Technology: 1991)

\bibitem{sw95}
J.K. Salmon and M.S. Warren, {\it "Skeletons from the Treecode closet"},  
{\it J. Comp. Phys.} {\bfseries 111} (1995) 136

\bibitem{sdg94}
L. Stiller, L.L. Daemene and J.E. Gubernatis, {\it J. Comp. Phys. } {\bfseries 115} (1994)
 550

\bibitem{ws95}
M.S. Warren and J.K. Salmon, {\it "A portable Parallel N-body code"},  
{\it unpublished }, (California Institute of Technology: 1995)

\bibitem{zqsw94}
W.H. Zurek, P.J. Quinn, J.K. salmon and M.S. Warren, {\it Astrophys. J} {\bfseries 431}
(1994)  559

\end{thebibliography}
\end{document}